\documentclass[reprint,showpacs,preprintnumbers,amsmath,amssymb,floatfix,prc]{revtex4}
\usepackage{graphicx}
\usepackage{dcolumn}
\usepackage{bm}

\def\beq{\begin{equation}}
\def\eeq{\end{equation}}
\def\be{\begin{eqnarray}}
\def\ee{\end{eqnarray}}

\newcommand{\lsim}{
 \mathrel{\setbox0=\hbox{$<$}\raise0.6ex\copy0\kern-\wd0
 \lower0.65ex\hbox{$\sim$}}}

\newcommand{\gsim}{
 \mathrel{\setbox0=\hbox{$>$}\raise0.6ex\copy0\kern-\wd0
 \lower0.65ex\hbox{$\sim$}}}

\begin{document}
\title{Nuclear effects in photoproduction of heavy quarks and vector mesons
    in ultraperipheral PbPb and pPb collisions at energies available at
    the CERN Large Hadron Collider}
\author{Adeola Adeluyi}
\author{C.A. Bertulani}
\affiliation{Department of Physics \& Astronomy,
Texas A\&M University-Commerce, Commerce, TX 75428, USA}
\author{M.J. Murray}
\affiliation{Department of Physics \& Astronomy, University of Kansas,
Lawrence, KS 66045-7582}

\date{\today}
\begin{abstract}
The comparison of photoproduction cross sections for $c\bar{c}$ 
and $b\bar{b}$ in PbPb and pPb collisions can give sensitivity to nuclear 
shadowing effects. The photoproduction of vector mesons is even more 
sensitive to the underlying gluon distributions.
In this study we present the cross sections and rapidity dependence of the  
photoproduction of heavy quarks and exclusive production of vector
mesons in ultraperipheral pPb and PbPb
collisions at the Large Hadron Collider at $\sqrt{s_{_{NN}}}=5$ TeV and
$\sqrt{s_{_{NN}}}=2.76$ TeV, respectively. The potentials of using
these processes for constraining nuclear gluon shadowing are
explored. It is found that photoproduction of $J/\psi$ and $\Upsilon$ in PbPb
collisions in particular exhibit very good sensitivity to gluon shadowing.
\end{abstract}
\maketitle
\vspace{1cm}
%
%

In view of recent and upcoming experiments at the Large Hadron
Collider (LHC) at CERN, we consider the inclusive photoproduction of 
heavy quarks ($c\bar{c}$ and $b\bar{b}$) and the 
exclusive elastic production of vector mesons ($J/\psi$ and 
$\Upsilon(1s)$) in ultraperipheral PbPb and pPb  collisions  
at $\sqrt{s_{_{NN}}}=2.76$ TeV and $5$ TeV respectively. The
theoretical framework used for our calculations has been described in 
two recent publications \cite{Adeluyi:2011rt,Adeluyi:2012ph}. The
method consists in using the strong electromagnetic fields generated
in relativistic heavy ion collisions \cite{BB88,BKN05} to produce
particles in $\gamma$-nucleus interactions. The large energies of the  
virtual photons produced in such reactions has stimulated many 
studies of the photoproduction of heavy mesons.
Previously, we have investigated the sensitivity of 
photoproduction of heavy quarks and exclusive production of vector
mesons at higher LHC energies to varying severity of gluon 
modifications \cite{Adeluyi:2011rt,Adeluyi:2012ph}. 
This idea, originally proposed in Ref. \cite{Goncalves:2001vs}, can be
used to constrain parton distribution functions from data on
photoproduction of heavy quarks and of vector mesons.

All four of the large LHC experiments, ALICE, ATLAS, CMS and LHCb,
have the capability to measure $J/\psi$ and $\Upsilon$,    
\cite{Abelev:2011md,Aad:2011sp,Aad:2011xv,Khachatryan:2010yr,
Chatrchyan:2011pe,Aaij:2011jh,LHCb:2012aa}. So far most LHC
measurements have been in the muon channel.  The calculations
presented in the present paper are reported in terms of rapidity 
distributions $d\sigma/dy$ in the center of mass frame. For PbPb
collisions at the LHC this is the same as the laboratory frame but for
pPb collisions the proton and lead beams have the same magnetic
rigidity which implies that the center of mass is shifted with respect 
to the laboratory frame by $\delta y = 0.5\ln(208/82) = 0.47 $. This 
rapidity shift presents both a challenge and an opportunity to the experiments. 
The acceptance of both CMS and ATLAS is symmetric about $y_{lab} = 0$
and extends up to $y_{lab} \approx 2.5$. The strong magnetic fields
and large amount of material before the muon stations imply that only
muons with a minimum momenta of order  5 GeV/c fire the trigger and
this causes both ATLAS and CMS to only reconstruct $J/\psi$ with a
minimum transverse momentum, $p_T$, near  $y_{lab} = 0$. However, 
at $y_{lab} \approx 2.5$ both experiments can reconstruct $J/\psi$s
down to $p_T = 0$. The extra mass of the $\Upsilon$ means that both
CMS and ATLAS have acceptance down to  $p_T = 0$ over their whole
rapidity range. Both ALICE and LHCb have asymmetric muon detectors
with acceptances down to $p_T=0$ for $2.5 \le y_{lab} \le 4.0$ [ALICE] 
$2 \le y_{lab} \le 4.5$ [LHCb]. If it were possible for these
collaborations to take both p+Pb and Pb+p data then almost the complete
phase space for ultraperipheral quarkonia production could be mapped out.

In the present work we follow closely the approach presented in Ref.
\cite{Adeluyi:2012ph}. All input parameters (except those related to
energy), computational details, and references therein are applicable
to the current study. We use the same sets of nuclear 
and photon parton distributions: MSTW08 \cite{Martin:2009iq}, EPS08 
\cite{Eskola:2008ca}, EPS09 \cite{Eskola:2009uj}, and 
HKN07 \cite{Hirai:2007sx} for nucleon/nuclear parton 
distributions and GRV \cite{Gluck:1991jc}, SaS1D \cite{Schuler:1995fk}
, and CJK2 \cite{Cornet:2004ng} for photon parton distributions. The 
characteristics of these distributions, especially the disparities in
the strength of the nuclear modifications of their gluon content, has
been treated in detail in \cite{Adeluyi:2012ph}.
Since in general the features and trends of the results obtained in the
previous study are faithfully reproduced in the present work, the
discussions and comments presented therein are also valid and the 
reader is therefore referred to \cite{Adeluyi:2012ph} for detailed exposition.
We will focus mainly on presenting the results, and hence in this 
respect the present report serves to augment and supplement the 
previous study contained in \cite{Adeluyi:2012ph}.

Let us first consider photoproduction of heavy quarks in  
ultraperipheral pPb and PbPb collisions where two rather different 
production mechanisms (direct and resolved) are present. In the direct  
mechanism the incident photon interacts directly with the target
(nucleus or proton) whereas in the resolved mechanism the incident
photon first fluctuates into a quark-antiquark pair (or even more
complicated partonic configuration) which then subsequently interacts 
hadronically with the target. At leading order the direct production involves
only the gluon distributions in the target while the resolved production
requires the distributions of light quarks and 
gluons in both photon and target. The total
production cross sections and rapidity distributions are 
of course the sum of the contributions from both processes. As in 
\cite{Adeluyi:2012ph} we will present the cross sections for three
nuclear parton distributions (MSTW08, EPS08, and EPS09) and the three 
photon parton distributions listed above. Also, for convenience, only the 
rapidity distributions using the GRV photon distributions are presented.

\begin{figure}[ht]
\includegraphics[width=\columnwidth]{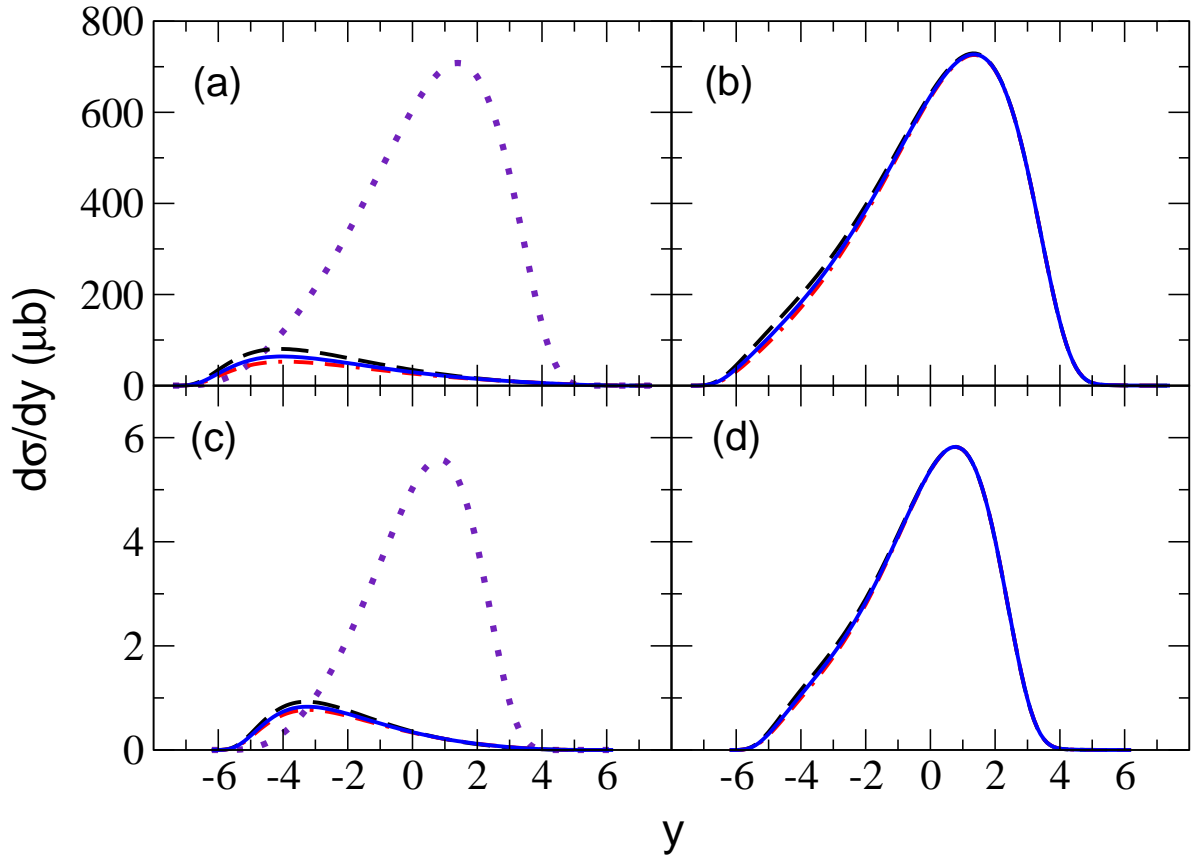}
\caption[...]{\label{fig:pphqpArapcc} (Color online) Rapidity
distributions of $c\bar{c}$ (top) and $b\bar{b}$ (bottom)
photoproduction in pPb collisions at $\sqrt{s_{_{NN}}}=5.0$ TeV using
the GRV photon parton distributions. The left hand panels (a and c) show 
the $\gamma$p and $\gamma$Pb contributions separately while the right hand 
panels (b and d) show the sum. Dotted line depicts 
the $\gamma$p contribution while the dashed (MSTW08), solid 
(EPS09), and dash-dotted (EPS08) lines correspond to $\gamma$Pb 
contributions with no shadowing, moderate, and strong
shadowing respectively.}
\end{figure}

In Table~\ref{ccbbpA} we present the direct and resolved cross 
sections for both $\gamma$p and $\gamma$Pb components of $c\bar{c}$ 
and $b\bar{b}$ production in pPb collisions. The respective $\gamma$p
and $\gamma$Pb rapidity distributions and their sums are 
displayed in Fig.~\ref{fig:pphqpArapcc}.
\begin{table}[!ht]
\caption{\label{ccbbpA} Cross sections (in $\mu$b) for photoproduction of 
$c\bar{c}$ and $b\bar{b}$ in ultraperipheral pPb collisions at $\sqrt{s_{_{NN}}}=5.0$ TeV.}
\begin{tabular}[c]{|c|c|c|c|c|c|c|}
\hline
\multicolumn{2}{|c|}{}&PDF            & Direct    & \multicolumn{3}{c|}{Resolved}  \\ 
\cline{5-7}
\multicolumn{2}{|c|}{}&                    &             & SaS1d   & GRV      & CJK   \\
\hline
& $\gamma$p &MSTW08          & 3250      & 323   & 534   & 665 \\
\cline{2-7}
 $c\bar{c}$ &                    &MSTW08          & 362       & 58   & 98    & 118   \\
                  &$\gamma$Pb &EPS08           & 238       & 50    & 85    & 101  \\
                  &                    &EPS09	    & 288      & 53   & 91    & 108  \\ \hline
\hline
                   & $\gamma$p &MSTW08        & 18.0      & 3.32   & 4.56   & 5.61  \\ \cline{2-7}
 $b\bar{b}$ &                     &MSTW08          & 2.80       & 0.90   & 1.29    & 1.58  \\
                  & $\gamma$Pb &EPS08          & 2.30       & 0.87    & 1.26   & 1.54  \\
                  &                     &EPS09	   & 2.50      & 0.88   &1.28   & 1.56  \\
\hline
\hline
\end{tabular} 
\end{table}
For both $c\bar{c}$ and $b\bar{b}$ production the $\gamma$p component is 
dominant due to the larger photon flux from the nucleus, leading to
asymmetric rapidity distributions skewed towards
the right in line with the convention adopted in
\cite{Adeluyi:2012ph}. While the influence of nuclear shadowing is 
somewhat discernible in the $\gamma$Pb component of $c\bar{c}$
photoproduction, this effect is minimised when one considers total
rapidity distributions due to the smallness of the $\gamma$Pb
component relative to the dominant $\gamma$p component. In view of
this, $c\bar{c}$ photoproduction in pPb collisions could potentially
serve as a good normaliser for the equivalent $c\bar{c}$
photoproduction in PbPb collisions when effects due to differences in
photon flux are taken into account.  
As evident from both Table~\ref{ccbbpA} and Fig.~\ref{fig:pphqpArapcc}, 
nuclear effects are rather unimportant in $b\bar{b}$ photoproduction,
since all three $\gamma$Pb rapidity distributions practically overlap.
On the other hand the resolved component is relatively more significant 
than in $c\bar{c}$ production. This attribute, in
conjunction with negligible sensitivity to nuclear effects and enhanced
sensitivity to  photon parton distributions as evinced in
Table~\ref{ccbbpA} suggests the potential for $b\bar{b}$
photoproduction to be of some use in constraining photon parton distributions.

\begin{figure}[!ht]
\includegraphics[width=\columnwidth]{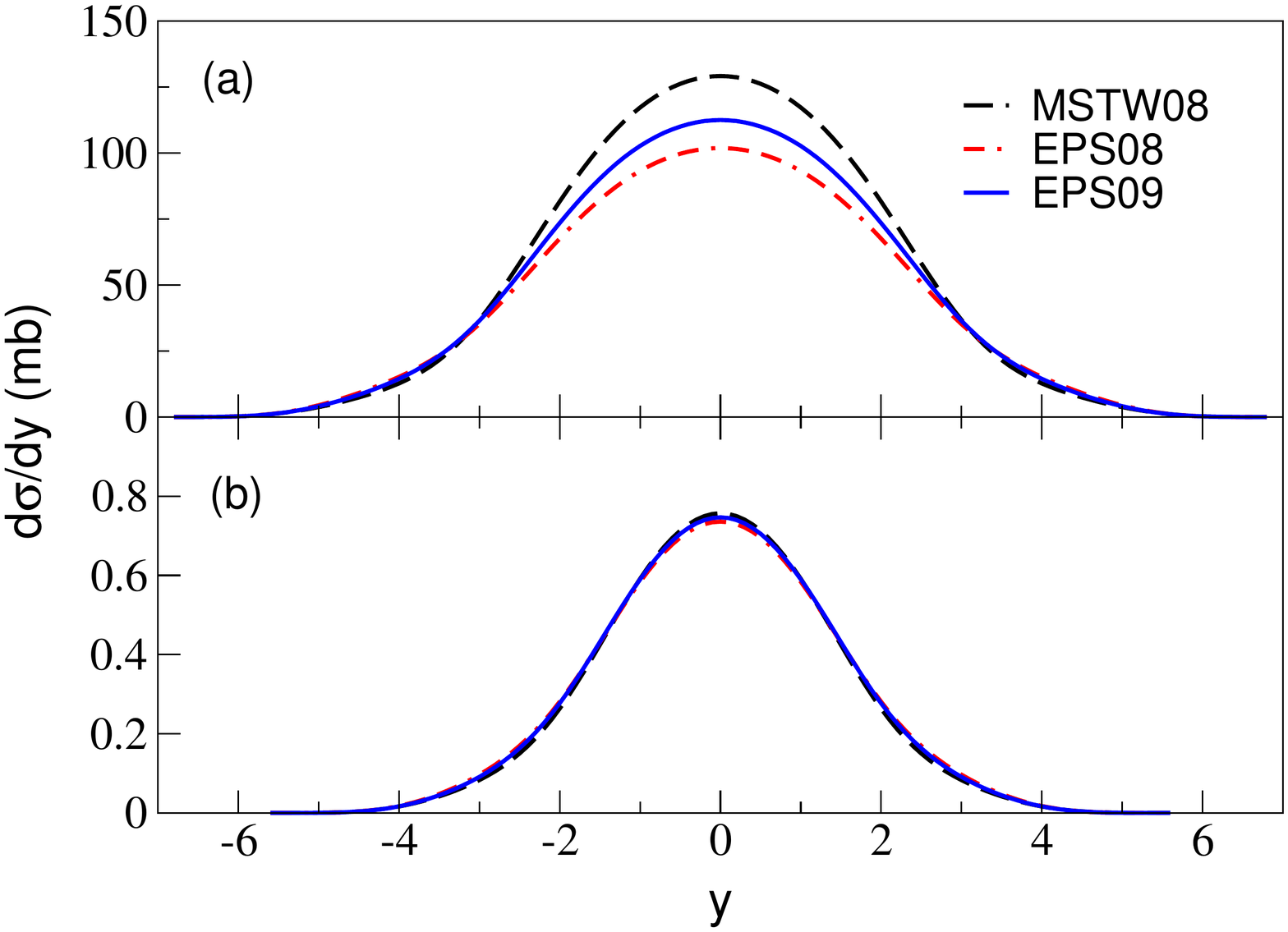}
\caption[...]{\label{thqrapcc} (Color online) Rapidity distributions of the
photoproduction of (a) $c\bar{c}$ (top) and (b) $b\bar{b}$ (bottom)
in PbPb collisions at $\sqrt{s_{_{NN}}}=2.76$ TeV using 
the GRV photon parton distributions. Dashed
line depicts result using the MSTW08 parton distributions (no nuclear
modifications). Solid and dash-dotted lines are results
from nuclear-modified parton distributions from EPS09 and EPS08 respectively.}
\end{figure}

The corresponding cross sections and rapidity distributions for 
PbPb collisions are shown in Table~\ref{ccbbAA} and Fig.~\ref{thqrapcc}. 
Here the participants are identical and each Pb nucleus can act as
both source and target of photons. Consequently the photon flux is the 
same and the rapidity distributions are symmetric about midrapidity
($y = 0$).
\begin{table}[!ht]
\caption{\label{ccbbAA} Cross sections for photoproduction of 
$c\bar{c}$ (in mb) and $b\bar{b}$ (in $\mu$b) in ultraperipheral 
PbPb collisions at the LHC.}
\begin{tabular}[c]{|l|c| c c c| c c c c c c |}
\hline
 &     PDF            && Direct              &&&&& Resolved && \\ 
\cline{6-11}
 &               &&             &&& SaS1d   && GRV      && CJK   \\
\hline 
 &      MSTW08          && 570.1     &&& 39.5   && 64.1    && 81.3  \\
$c\bar{c}$ (mb)&  EPS08           && 469.8       &&& 39.4    && 64.1    && 81.0  \\
&          EPS09	        && 511.1      &&& 39.6   && 64.5    && 81.6  \\
\hline
&       MSTW08          && 2277.0      &&& 293.6   && 388.5    && 467.8  \\
$b\bar{b}$ ($\mu$b)& EPS08           && 2280.0       &&& 310.5    && 413.5    && 501.1  \\
&       EPS09	        && 2291.8      &&& 304.2   && 404.3    && 489.1  \\
\hline
\end{tabular} 
\end{table}
For $c\bar{c}$ production the effect of shadowing is quite appreciable 
especially in the rapidity interval $-2 < y < 2$, and thus this interval offers 
good constraining potential. The effect of antishadowing is
discernible in the intervals $3 < y < 5$ and $-5 < y < -3$, although
the magnitude is rather small. As in pPb collisions, $b\bar{b}$
production shows little sensitivity to nuclear effects, with the resolved
component being more significant.

\begin{figure}[!ht]
\includegraphics[width=\columnwidth]{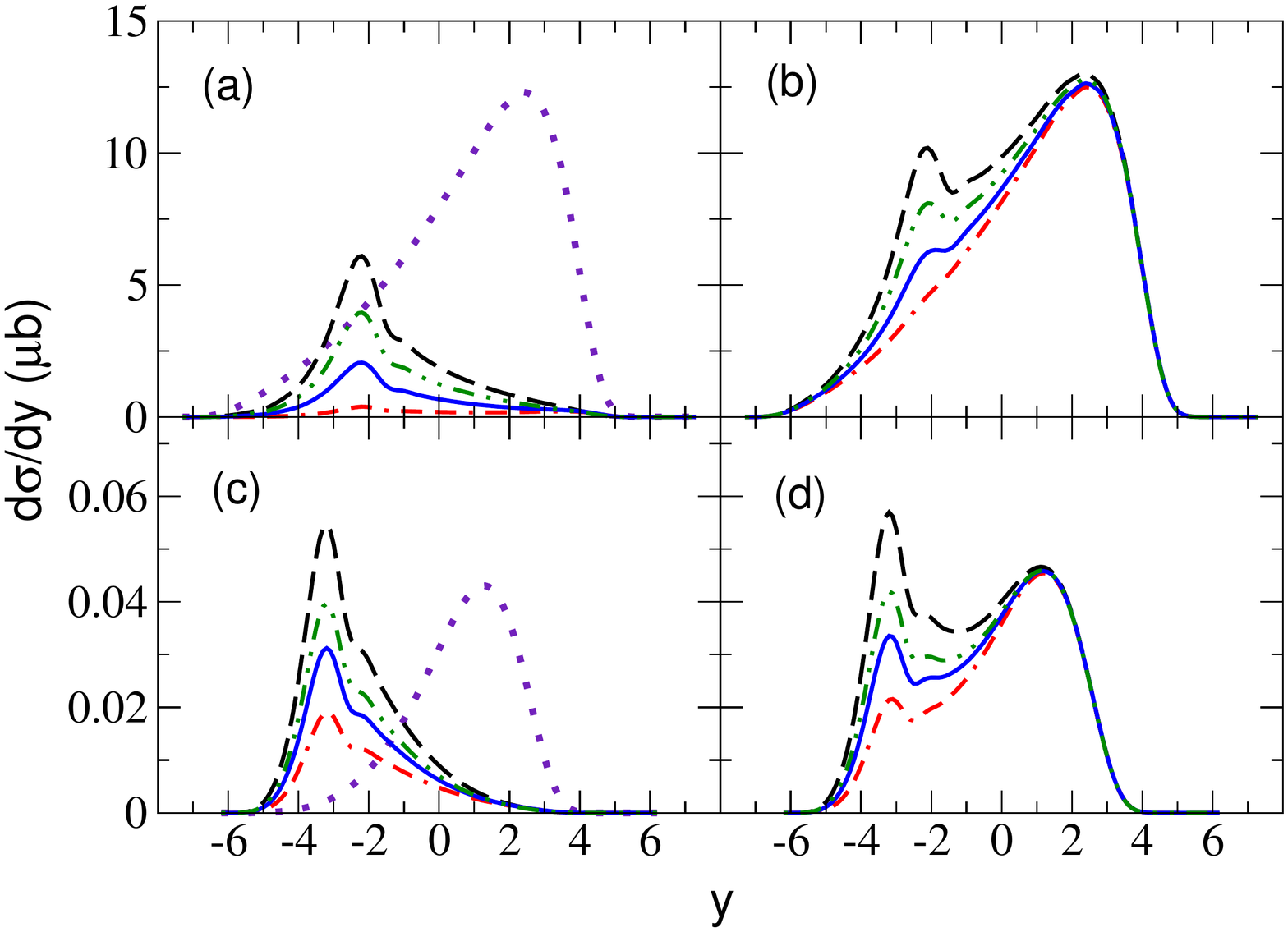}
\caption[...]{\label{jpsiupsipPb} (Color online) Rapidity distributions of 
exclusive photoproduction of $J/\psi$ (top) and  $\Upsilon$ (bottom)
in pPb collisions at $\sqrt{s_{_{NN}}}=5$ TeV.  
The left hand panels (a and c) show the $\gamma$p and $\gamma$Pb
contributions separately while the right hand panels (b and d) show the sum. 
 Dotted line depicts 
the $\gamma$p contribution while the dashed (MSTW08), dash-double-
dotted (HKN07), solid 
(EPS09), and dash-dotted (EPS08) lines correspond to $\gamma$Pb 
contributions with no shadowing, weak, moderate, 
and strong shadowing respectively.}
\end{figure}

We now present the results on elastic photoproduction of the $J/\psi$ 
and $\Upsilon(1s)$. As discussed in \cite{Adeluyi:2011rt,Adeluyi:2012ph} 
the production mechanism for these vector mesons involves
the square of the nuclear/nucleon gluon distribution. This quadratic
dependence leads to a dramatic increase in the sensitivity of both 
cross sections and rapidity distributions to nuclear effects
(predominantly shadowing) on gluon distributions.  
In Table~\ref{tjpsiupsipPb} we present the component 
and total cross sections for the elastic photoproduction of $J/\psi$ 
and $\Upsilon$ in ultraperipheral pPb collisions at the LHC. The 
associated rapidity distributions are shown in Fig.~\ref{jpsiupsipPb}.
\begin{table}[!ht]
\caption{\label{tjpsiupsipPb} Cross sections for elastic photoproduction of 
$J/\Psi$ (in $\mu$b) and $\Upsilon$ (in nb) in ultraperipheral 
pPb collisions.}
\begin{tabular}[c]{|l|c|ccc c c c c c c c| c c c c}
\hline
&            PDF &             && $\gamma$p    &&& $\gamma$Pb    &&& Total & \\
\hline
&            MSTW08&           && 63.6        &&& 18.3         &&& 81.9 & \\
$J/\Psi$&    EPS08&            &&              &&& 1.8          &&& 65.4 & \\		       
($\mu$b)&    EPS09&	         &&              &&& 6.6        &&& 70.2  & \\		      
&            HKN07&            &&              &&& 12.0         &&& 75.6  & \\	
\hline
&            MSTW08&           && 149.6        &&& 137.0         &&& 286.6 & \\
$\Upsilon$&  EPS08&            &&              &&& 54.8          &&& 204.4 & \\		       
(nb)&        EPS09&	       &&              &&& 82.9        &&& 232.5  & \\		      
&            HKN07&            &&              &&& 101.5         &&& 251.1  & \\	
\hline                              
\end{tabular}
\end{table}
%

Unlike photoproduction of $c\bar{c}$ and $b\bar{b}$
in pPb collisions which is practically insensitive to
shadowing, the enhanced sensitivity to gluon shadowing due to the
quadratic dependence is already apparent here. Thus even though the $\gamma$p
contribution is dominant in the case of $J/\psi$ production , the
$\gamma$Pb contributions from both MSTW08 (no shadowing) and HKN07
(weak shadowing) are still relatively 
appreciable. On the other hand the severity of the gluon 
shadowing present in EPS08 is enough to render its $\gamma$Pb
contribution almost negligible. For $\Upsilon(1s)$ production the 
$\gamma$Pb component contributes significantly, and is in fact
comparable to the $\gamma$p contribution in the case of MSTW08.
Due to this, the effect of gluon shadowing is more clearly reflected
in the total rapidity distributions and thus 
$\Upsilon(1s)$ production may potentially be of some use in
constraining gluon shadowing, especially in the $-4 < y < -1$ rapidity
interval. 

\begin{figure}[!ht]
\includegraphics[width=\columnwidth]{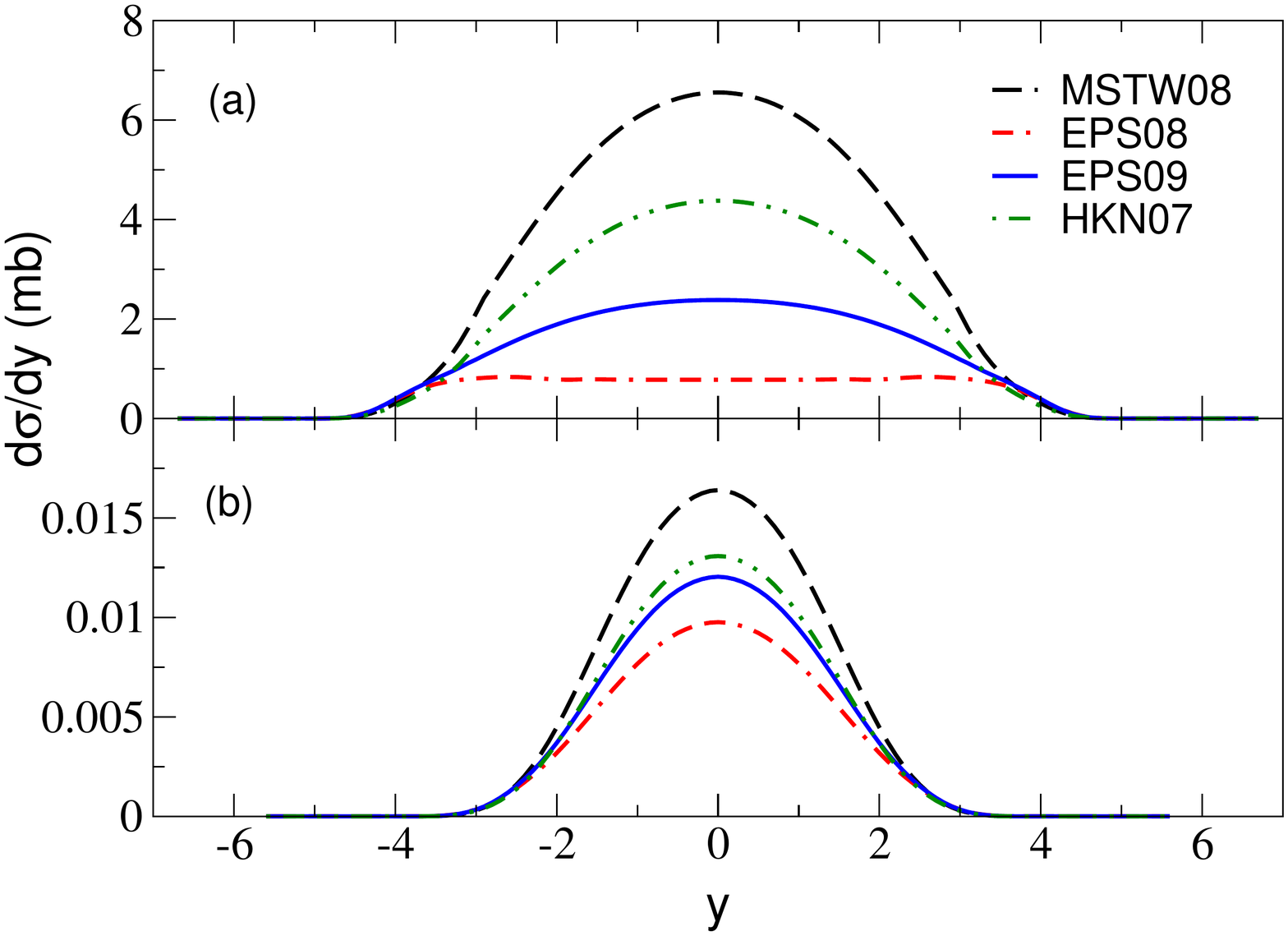}
\caption[...]{\label{jpsiupsiPbPb} (Color online) Rapidity distributions of 
exclusive photoproduction of (a) $J/\psi$ (top) and (b) $\Upsilon(1s)$ (bottom) in PbPb collisions at $\sqrt{s_{_{NN}}}=5.0$ TeV. 
Dashed, solid, dash-dotted, and dash-double-dotted lines are results
from MSTW08, EPS09, EPS08, and HKN07 parton distributions respectively.}
\end{figure}

The results on $J/\psi$ and $\Upsilon(1s)$ production in
ultraperipheral PbPb collisions are presented in
Table~\ref{tjpsiupsiPbPb} and in
Fig.~\ref{jpsiupsiPbPb}. The differences in 
the predicted cross sections and rapidity distributions are markedly
clear cut, especially for $J/\psi$. Thus photoproduction of $J/\psi$
and $\Upsilon(1s)$ in ultraperipheral PbPb collisions 
can serve as an excellent probe of gluon shadowing 
as well as a good discriminator of the different gluon shadowing sets
utilized in the current study.
\begin{table}[!ht]
\caption{\label{tjpsiupsiPbPb} Total cross sections for elastic photoproduction of 
$J/\psi$ (in mb) and $\Upsilon(1s)$ (in $\mu$b) in ultraperipheral 
PbPb collisions.}
\begin{tabular}[c]{|l|c c c|}
\hline
PDF                    && $J/\Psi$ (mb) & $\Upsilon$ ($\mu$b)  \\   
\hline
MSTW08              && 32.6         & 51.3  \\		      
EPS08               && 6.3          & 32.2  \\		       
EPS09	            && 13.9         & 39.0  \\		      
HKN07               && 22.1         & 41.3  \\
\hline	               
\end{tabular}
\end{table}

Let us briefly compare the present results to those
at higher collision energies (pPb at  $\sqrt{s_{_{NN}}}=8.8$ TeV and 
PbPb at $\sqrt{s_{_{NN}}}=5.5$ TeV) presented in \cite{Adeluyi:2012ph}.
For pPb collisions the cross sections for $c\bar{c}$ ($b\bar{b}$) production at 
$\sqrt{s_{_{NN}}}=8.8$ TeV are approximately $1.8$ ($2.1$) times those
at $\sqrt{s_{_{NN}}}=5.0$ TeV. The relative $\gamma$Pb
contributions are almost equal and nuclear effects are practically the 
same at both energies for both heavy quarks. 
For PbPb the cross sections at $\sqrt{s_{_{NN}}}=5.5$
TeV are approximately $2.1$ ($2.8$) times those at $\sqrt{s_{_{NN}}}=2.76$ TeV
for $c\bar{c}$ ($b\bar{b}$). Nuclear effects are about $27\%$ larger
for $c\bar{c}$ although shadowing trends are identical. 
The case of $b\bar{b}$ is interesting: strong influence of
antishadowing results in both  EPS08 and EPS09 $b\bar{b}$ cross
sections at $\sqrt{s_{_{NN}}}=2.76$ TeV being larger than that of MSTW08. 
This contrast with the behavior at $\sqrt{s_{_{NN}}}=5.5$ TeV where the
usual shadowing trend prevails. 
 
Further differences manifest in photoproduction of $J/\psi$ and 
$\Upsilon(1s)$ in pPb collisions. For  $J/\psi$ the relative 
$\gamma$Pb contributions at $\sqrt{s_{_{NN}}}=5.0$ TeV are about twice those at
$\sqrt{s_{_{NN}}}=8.8$ TeV  and about $30\%$ larger for
$\Upsilon(1s)$. Shadowing effects are therefore more pronounced at 
lower energy and consequently better suited for constraining purposes.
As expected total cross sections for $J/\psi$ ($\Upsilon(1s)$) 
are approximately a factor of $2.5$ ($2.2$) 
larger than at $\sqrt{s_{_{NN}}}=8.8$ TeV. For PbPb collisions although
the cross sections are larger at $\sqrt{s_{_{NN}}}=5.5$ TeV, shadowing 
effects are almost the same (for $J/\psi$) or slightly larger 
(for $\Upsilon(1s)$) than at $\sqrt{s_{_{NN}}}=2.76$ TeV. Thus the  
constraining abilities at both energies are almost at par.

In conclusion we offer the following remarks: the dependence on parton
distributions is linear in photoproduction of heavy quarks  and
different modifications are superimposed due to the
integration over the momentum fraction $x$. Despite these limitations, 
both cross sections and rapidity distributions for $c\bar{c}$ in PbPb
collisions manifest appreciable sensitivity to
shadowing around midrapidity and very slight sensitivity to antishadowing 
at more forward and backward rapidities. Thus $c\bar{c}$
photoproduction offers good constraining potential for shadowing. Both $c\bar{c}$ and 
$b\bar{b}$ total photoproduction cross sections and rapidity
distributions in pPb collisions show little sensitivity
to nuclear modifications. The resolved components are however
appreciable, especially for $b\bar{b}$, and are significantly dependent on
the choice of photon parton distributions. Thus it seems feasible that they could be
of some use in constraining photon parton distributions.  
The outlook for constraining gluon shadowing is better still in 
the case of vector meson production.   
Here the quadratic dependence on gluon modifications makes elastic
photoproduction of vector mesons particularly attractive for 
constraining purposes. In particular both the cross sections and rapidity 
distributions for $J/\psi$ and $\Upsilon(1s)$ photoproduction 
in PbPb collisions exhibit very good sensitivity to gluon shadowing.

\bigskip

We acknowledge support by the US Department of Energy grant  DE-FG02-08ER41533  and the Research Corporation.
%

\end{document}